\documentclass[%
reprint,
aps,
prb,
superscriptaddress,
]{revtex4-1}

\usepackage[dvipdfmx]{hyperref}
\hypersetup{colorlinks=true, linkcolor=blue, urlcolor=blue, citecolor=blue}
\usepackage[dvipdfmx]{graphicx}
\usepackage{amsmath} \usepackage{amssymb}
\usepackage{accents} \usepackage{mathrsfs} \usepackage{setspace}
\usepackage{color}
\definecolor{dgreen}{RGB}{00, 120, 00} \definecolor{dblue}{RGB}{00, 00, 220}
\definecolor{lgreen}{RGB}{46, 139, 87} 
\usepackage{ulem}

\newcommand{\bs}[1]{\boldsymbol{#1}} 
\newcommand{\tr}[1]{\mathrm{Tr}\left[#1\right]}

\newcommand{\ut}[1]{\undertilde{#1}} 
 \newcommand{\vphi}[0]{\varphi}

\begin{document}

%
%


\title{Destruction of surface states of 
($d_{zx}+id_{yz}$)-wave superconductor\\ 
by surface roughness: application to Sr$_2$RuO$_4$}

\author{Shu-Ichiro Suzuki}
\email{s.suzuki-1@utwente.nl}
\affiliation{MESA+ Institute for Nanotechnology, University of Twente, 
7500 AE Enschede, The Netherlands}

\author{Satoshi Ikegaya}
\affiliation{Department of Applied Physics, Nagoya University, 
Nagoya 464-8603, Japan}

\author{Alexander A. Golubov}
\affiliation{MESA+ Institute for Nanotechnology, University of Twente, 
7500 AE Enschede, The Netherlands}

\date{\today}

\begin{abstract}
	The fragility of the chiral surface current of ($d_{zx}+id_{yz}$)-wave
	superconductor, a potential candidate for Sr$_2$RuO$_4$, against surface roughness is demonstrated utilizing the
	quasiclassical Eilenberger theory. Comparing the chiral surface
	currents of ($d_{zx}+id_{yz}$)-wave and ($p_{x}+ip_{y}$)-wave
	pairings, we conclude the chiral current for ($d_{zx}+id_{yz}$)-wave
	SC is much more fragile than that for the ($p_x+ip_y$)-wave one. The
		difference can be understood in terms of the orbital symmetry of
		the odd-frequency Cooper pairs arising at the surface. 
	Our results show the ($d_{zx}+id_{yz}$)-wave scenario can explain the null
	spontaneous magnetization in Sr$_2$RuO$_4$ experiments. 
\end{abstract}

\pacs{pacs}

\thispagestyle{empty}

\maketitle

\textit{Introduction}.---%
The determination of the pairing symmetry in Sr$_2$RuO$_4$ (SRO)
superconductors (SCs) has been an unsolved problem for more than a
quarter century~\cite{maeno_94,maeno_03,kallin_09,maeno_17}.  In
the last few years, nevertheless, researchers in this field undergo a
remarkable paradigms shift.  Specifically, recent precise experiments
on spin susceptibility~\cite{brown_19,ishida_20,brown_20,hayden_20}
appear to contradict a spin-triplet odd-parity superconducting state
with broken time-reversal symmetry (TRS)\cite{Rice}, which had heretofore been
the leading candidate in SRO.  Alternatively, an exotic
\textit{inter-orbital-singlet} spin-triplet even-parity state with
broken time-reversal symmetry has come under the
spotlight~\cite{agterberg_20(r),fukaya_22} because it can explain
recent two remarkable experimental observations, i.e., a sharp jump in
the shear elastic constant $c_{66}$ at the superconducting transition
temperature measured by ultrasound
experiments~\cite{proust_20,ramshaw_20}, and a stress-induced split
between the onset temperatures for the superconducting state and
broken TRS state measured by muon spin-relaxation
experiments~\cite{grinenko_20,grinenko_21}.  Nowadays, careful and
intensive verification for the realization of the inter-orbital
superconducting state in SRO has been underway.

On the basis of a microscopic model for the inter-orbital
superconducting state of SRO~\cite{agterberg_20(r)}, the
superconducting gap on the three Fermi surfaces of SRO has a
($d_{zx}+id_{yz}$)-wave pairing symmetry (i.e., $d+id'$-wave SC).  It has been
shown that the $d+id'$-wave SC hosts characteristic surface
states~\cite{kobayashi_15,tamura_17,suzuki_20,ikegaya_21(3)}.  At
material surfaces parallel to the $z$-axis (i.e. the $c$-axis of the
SRO), the $d+id'$-wave  SC exhibits dispersing chiral surface states
due to the chiral pairing symmetry with fixed $k_z$.  Moreover, the
pure odd-parity nature with respect to $k_z$ results in the emergence
of dispersion-less zero-energy surface states at the surfaces
perpendicular to the $z$-axis.  Thus, observations of these
surface states can be the conclusive evidence for the inter-band
superconducting state in SRO.  However, scanning superconducting
quantum interference devise experiments\cite{moler_05, nelson_07} have
not detected the expected spontaneous edge current due to the chiral
surface states~\cite{Matsumoto_JPSJ_1999, Furusaki_PRB_2001, Stone_04,
Nagato_11, Bakurskiy_14, suzuki_16}, and tunneling spectroscopy
measurements along the $z$-axis did not observe a zero-bias
conductance peak suggesting the dispersion-less
zero-energy surface states~\cite{flouquet_09,
kivelson_13, madhavan_20}. Therefore, when we take the experimental
observations at face value, the inter-orbital superconducting state
with a $d+id'$-wave superconducting gap seems to be excluded.

\begin{figure}[b]
	\includegraphics[width=0.46\textwidth]{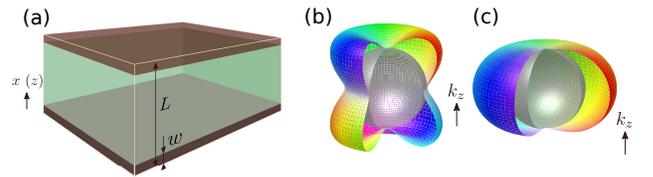}
	\caption{(a) Schematics of the system. The surfaces are parallel to
	the either of $x$ or $z$ axis. The widths of the superconductor and
	disordered regions are denoted by $L$ and $w$. The translational symmetry is assumed 
	in the direction parallel to the surfaces. The pair potentials
	of ($d_{zx}+id_{yz}$)- and ($p_x+ip_y$)-wave superconductors are shown
	in (b) and (c). The colour indicates $\mathrm{arg}[\Delta(k)]$. The inner silver sphere represents the
	Fermi sphere. }
\label{fig:sche}
\end{figure}

In this Letter, we study the influence of surface roughness on the
surface states of the $d+id'$-wave  SC.  
The most straightforward numerical simulation is adding random
potentials to the microscopic three-orbital
Hamiltonian\cite{agterberg_20(r)}. However, such numerical
simulation requires significantly large systems in real-space, 
ensemble averaged of impurity configurations, and 
self-consistent treatments for the order parameter, meaning that 
it would be impossible to implement owing to the
prohibitive numerical costs.  Alternatively, we employ the
quasiclassical Eilenberger theory for a
simple single-band and clarify essential properties of the surface
states of the $d+id'$-wave  SC.  As a result, we demonstrate
that the surface current due to the chiral surface states and the
sharp zero-energy peak in the surface density of states due to the
dispersion-less zero-energy surface states are easily destroyed by
surface roughness.  Importantly, the vulnerability of the surface
states is owing to a roughness-induced destructive interference effect
which is inevitable with the $d+id'$-wave pairing symmetry.
Namely, the surface states of the $d+id'$-wave  SC in the
presence of surface roughness are fragile regardless of details of the
model.  We will conclude that the absence of experimental signatures
from the surface states does not contradict with the inter-orbital
$d+id'$-wave superconducting states in SRO because surface roughness is inevitable
in real-life experiments.

\textit{Quasiclassical Eilenberger theory}.---
We examine the effects of surface roughness utilizing the quasiclassical Eilenberger theory \cite{Eilenberger}. 
The SC has a pair of parallel surfaces which are perpendicular to the $x$ or $z$ axis as shown in Fig.~\ref{fig:sche}(a). 
The thin dirty layers with the width $w$ are introduced. 
The Green's functions obey the
Eilenberger equation: 
\begin{align}
  & i \bs{v}_F \cdot \bs{\nabla} \check{g}
	+ \left[ \, i\omega_n \check{\tau}_3+\check{H},~\check{g} \right]_-
	= 0, 
	\label{eq:Eilen}
	\\[2mm]
  & \check {H}
  =\check {\Delta}+ \check {\Sigma}
	= \left( \begin{array}{rr}
	  \hat{\xi } &  
	  \hat{\eta} \\[1mm]
	  \hat{\ut{\eta}} & 
	  \hat{\ut{\xi }} \\
	\end{array} \right), 
	\hspace{4mm}
  \check {\Sigma}
  = \frac{i}{2 \tau_0} \langle \check {g} \rangle, 
	\\[2mm]
	& \check {g} = \left( \begin{array}{rr}
	 \hat{    g}  &  
	 \hat{    f}  \\[1mm]
	-\hat{\ut{f}} & 
	-\hat{\ut{g}} \\
	\end{array} \right), 
	\hspace{4mm}
	 \check {\Delta} = \left( \begin{array}{cc}
	 0&  
	 \hat{\Delta} \\[1mm]
	 \hat{\ut{\Delta}} & 
	 0\\
	\end{array} \right), 
	\hspace{4mm}
\end{align}
where 
$
  \langle \cdots \rangle
  = \int_0^\pi \int_{-\pi}^\pi \cdots 
	{\sin \theta d\vphi d\theta}/{4 \pi}, 
	$
$\check{g}= \check{g} (\bs{r},\bs{k},i\omega_n)$ is the quasiclassical
Green's function in the Mastubara representation, 
$\check {\Delta} = \check {\Delta}(\bs{r},\bs{k})$ is the pair-potential matrix, 
$\check {\Sigma} = \check {\Sigma}(\bs{r},i \omega_n)$ is the 
self-energies by the impurity scatterings, and we assume the system is in equilibrium. 
The mean free path is denoted $\ell = v_F \tau_0$ with $\tau_0$ being 
the mean free time 
that is fixed at a
certain value in the disordered region but infinitely large in the
other place. 
In this Letter, the accents
$\check{\cdot}$ and $\hat{\cdot}$ means matrices in particle-hole and spin space. 
The identity matrices in particle-hole and spin space 
are respectively denoted by $\check{\tau}_0$ and $\hat{\sigma}_0$. 
The Pauli matrices are denoted by $\check{\tau}_j$ and $\hat{\sigma}_j$ 
with $j \in {1, 2, 3}$. 
All of the functions satisfies
the symmetry relation $\hat{K} (\bs{r},\bs{k},i \omega_n) =
[\hat{\ut{K}} (\bs{r},-\bs{k},i \omega_n)]^*$, where the unit vector
$\bs{k}$ represents the direction of the Fermi momentum. Effects of the vector potential are ignored because it affects on surface states only quantitatively. 

The Eilenberger equation \eqref{eq:Eilen} can be simplified by the
so-called Riccati parameterization\cite{Schopohl_PRB_95, 
Eschrig_PRB_00, Eschrig_PRB_09}. The Green's function can be
expressed in terms of the coherence function $\hat{\gamma}= \hat{\gamma}(\bs{r}, \bs{k}, i \omega_n)$:
\begin{align}
	& \check {g} = 2 \left( \begin{array}{rr}
	 \hat{\mathcal{G}} &  
	 \hat{\mathcal{F}} \\[1mm]
	-\hat{\ut{\mathcal{F}}} & 
	-\hat{\ut{\mathcal{G}}} \\
	\end{array} \right) - \check{\tau}_3, 
	\\[1mm]
	& \hat{\mathcal{G}} = (1-\hat{\gamma} \hat{\ut{\gamma}} )^{-1},\hspace{6mm}
	  \hat{\mathcal{F}} = (1-\hat{\gamma} \hat{\ut{\gamma}})^{-1}\hat{\gamma}.
\label{eq:Ric-Para}
\end{align}
The equation for $\hat{\gamma}$ is given by 
\begin{align}
  & (i \bs{v}_F \cdot \bs{\nabla} + 2 i \omega_n ) \hat{\gamma}
	+ \hat{\xi} \hat{\gamma} - \hat{\gamma}\hat{\ut{\xi}}
	-\hat{{\eta}}+ \hat{\gamma} \hat{\ut{\eta}} \hat{\gamma}= 0. 
	\label{eq:Riccati03}
\end{align}
Assuming no spin-dependent potential and single-spin $\hat{\Delta}$, 
we can parameterize the spin structure of the functions: 
\begin{align}
	& \hat{\Delta} = i \Delta_{\bs{k}, \nu} (i \hat{\sigma}_\nu \hat{\sigma}_2), 
	\\
	& \hat{\ut{\Delta}} = -i \Delta_{-\bs{k}, \nu}^* (i \hat{\sigma}_\nu \hat{\sigma}_2)^*
	= i \Delta_{\bs{k}, \nu}^* (i \hat{\sigma}_\nu \hat{\sigma}_2)^\dagger
	\\
	&     \hat{g}  =     g         \hat{\sigma}_0, ~ ~
	      \hat{f}  =     f _\nu (i \hat{\sigma}_\nu \hat{\sigma}_2), ~ ~
		\ut{\hat{f}} = \ut{f}_\nu (i \hat{\sigma}_\nu \hat{\sigma}_2)^\dagger, 
		\\
  & \hat{\eta} = i \eta_{\nu} (i \hat{\sigma}_\nu \hat{\sigma}_2),
	~ ~ 
\ut{\hat{\eta}} = i \ut{\eta}_{\nu} (i \hat{\sigma}_\nu \hat{\sigma}_2)^\dagger, 
\end{align}
where $\nu = 0$ ($\nu \in \{1, 2, 3 \}$) is for the spin-singlet (spin-triplet) SC. In
the following, we make $\nu$ explicit only when necessary. 
Equation~\eqref{eq:Riccati03} can be reduced to 
\begin{align}
  & \bs{v}_F \cdot \bs{\nabla}  \gamma
	+ 2 \tilde{\omega} \gamma
	  -     \eta 
		+ \ut{\eta}
	\gamma^2= 0, 
	\label{eq:Riccati07}
	\\
	& \tilde{\omega} = 
	\omega_n 
	+\frac{\mathrm{Re} \langle g \rangle }{2 \tau_0}, 
	\\
	& \eta_{\nu} = \Delta_{\bs{k}} + \frac{\langle f \rangle}{2 \tau_0}, 	
	\hspace{6mm} \ut{\eta}_{\nu} = \Delta_{\bs{k}}^* -S_\nu
	\frac{\langle f \rangle^*}{2 \tau_0}. \label{eq:eta1}
\end{align}
The coherence functions in the homogeneous limit $\bar{\gamma}$ is given by
\begin{align}
  & 
  \bar{\gamma} (\bs{k}, i \omega_n) = 
	\frac{s_o \Delta_{\bs{k}} }
	{ |\omega_n| + \sqrt{\omega_n^2 + |\Delta_{\bs{k}}|^2} }, 
\end{align}
with $s_o=\mathrm{sgn}[\omega_n]$ and $\bar{\cdot}$ means the bulk
value. 

The momentum dependence of the pair potential is assumed as
\begin{align}
	& \Delta_{\bs{k}} 
	= \left\{ \begin{array}{cr}
	  2 (\Delta_1 k_x+i\Delta_2k_y) k_z & \text{for $d+id'$-wave,} \\
	     \Delta_1 k_x+i\Delta_2k_y      & \text{for $p+ip'$-wave,} \\
	\end{array} \right. 
\end{align}
where we put the factor $2$ in the $d+id'$-wave case such that 
$\mathrm{max}[\Delta_{\bs{k}}] = \bar{\Delta}$ in the homogeneous
limit. 
The schematic gap amplitudes in the bulk are shown in Fig.~\ref{fig:sche}(b) 
and \ref{fig:sche}(c), where the color means the phase of the pair
potential $\mathrm{arg}[\Delta(\bs{k})]$. 
The spatial dependence of the pair potentials are determined by the
self-consistent gap equation which relates $f$ and $\Delta$: 
\begin{align}
	& \Delta_\mu(\bs{r})
	=
	2 \lambda N_0 \frac{\pi}{i \beta} \sum_{\omega_n}^{\omega_c}
	\langle V_\mu(\bs{k}') f(\bs{r},\bs{k}',i \omega_n) \rangle, 
	\\
	& \lambda 
	= \frac{1}{2 N_0}
	\left[
	\ln\frac{T}{T_c} 
	+ \sum_{n=0}^{n_c}
	\frac{1}{n+1/2}
	\right]^{-1}, 
\end{align}
where $\mu=1$ or $2$, $\beta=1/T$, $T_c$ is the critical temperature, $N_0$ is the density of the states (DOS) in the normal state at the Fermi energy, and $n_c$ is the cutoff
integer.
The corresponding attractive potentials are 
$(V_1, V_2) = ({15}/{2}) (k_z k_x, k_y k_z)$ for the $d+id'$-wave
and 
$(V_1, V_2) = 3 (k_x, k_y)$ for the $p+ip'$-wave
SCs. 

The charge current, local DOS, and angle-resolved DOS are
calculated from the Green's function: 
\begin{align}
	&j_y(\bs{r})
	= 
	eN_0 \frac{\pi}{i \beta} \sum_{\omega_n}^{\omega_c}
	\langle k_y
	\tr{\check{\tau}_3 \check{g}(\bs{r},\bs{k},i \omega_n)}  \rangle, 
	\\
	&N(\bs{r}, E)
	= \int
	N_{\mathrm{AR}}(\bs{r}, k_y, E) dk_\parallel, 
	\\
	& \frac{N_{\mathrm{AR}}}{N_0}	= \sum_{\alpha = \pm 1}
	\mathrm{Re}[g(\bs{r}, \pm k_\perp, \bs{k}_\parallel, i \omega_n)]_{i
	\omega_n \to E+i\delta}, 
\end{align}
with $e <0$ is the charge of an electron, $\bs{k}_\parallel$
($k_\perp$) is the momentum parallel (perpendicular) to the surface. 
The amplitude of the subdominant Cooper pairs can be extracted from
the anomalous Green's function:
\begin{align}
  & f_{p_z} = \langle k_z f \rangle, 
	\hspace{6mm}
  f_{s  } = \langle f \rangle, 
  \label{}
\end{align}

In the numerical simulations, we fix the parameters: $L=80 \xi_0$, $w=3\xi_0$, $\omega_c = 10
\pi T_c$, $T=0.2T_c$, $\delta=0.01 \bar{\Delta}$ with $\xi_0 = \hbar
v_F / 2 \pi T_c$ being the coherence length. 

\begin{figure}[tb]
	\includegraphics[width=0.46\textwidth]{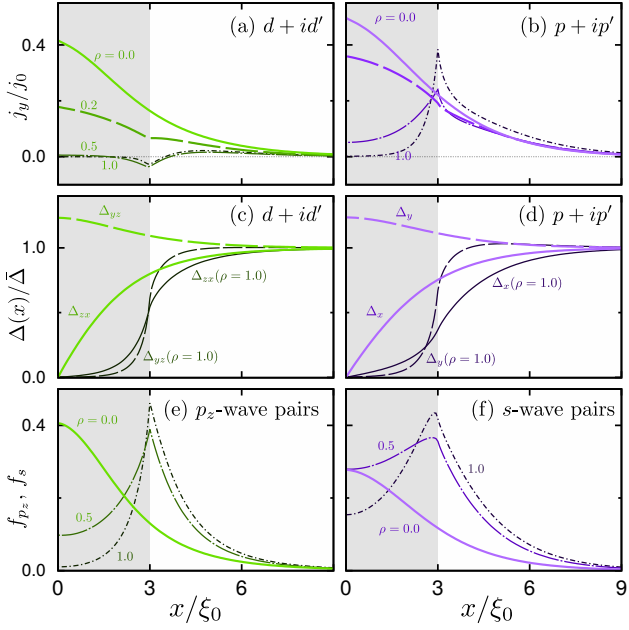}
	\caption{Calculated results for the (a)(c)(e) $d+id'$-wave and
	(b)(d)(e) $p+ip'$-wave SCs. (a)-(b) The edge-current density in the
	$y$ direction, (c)-(d) self-consistent pair potentials $\Delta$, 
	and (e)-(f) subdominant pair amplitudes are shown.
  The surface-roughness parameters are fixed as $\rho = \xi_0/\ell$ and $w=3\xi_0$. 
	The current density is normalized to $j_0 = |e|v_F N_0 \pi T_c$. }
  \label{fig:cur}
\end{figure}

\textit{Chiral surface current and pair functions}.---%
We first discuss the result for the open surface in the $x$-axis
direction.  The spatial profiles of $j_y$ and $\Delta$ are shown in
\ref{fig:cur}(a)-\ref{fig:cur}(d), where the results for the
$d+id'$-wave and $p+ip'$-wave SCs are shown in the left and right
panels respectively. Figures \ref{fig:cur}(a) and \ref{fig:cur}(b)
show the chiral surface current (CSC) for the $d+id'$-wave SC is much
more sensitive to the
surface roughness than the $p+ip'$-wave case. Even with a weak surface
roughness (i.e., $\xi_0/\ell = 0.5$), the CSC for the $d+id'$-wave SC is almost 
zero\footnote{We have confirmed the similar fragility of the
CSC in the $f+if'$-wave SC with $\Delta_{\bs{k}} \sim (\Delta_1
k_x+i\Delta_2k_y) (5k_z^2-1)$},
whereas that for the $p+ip'$-wave SC is sufficiently large to be
observed \cite{Bakurskiy_14, suzuki_16} where the peak in the current density
moves from the surface to the internal surface between the disordered
and ballistic regions. 
The pair potentials for the both SCs show qualitatively the
same behaviour to the surface roughness. At the clean surface, the
component that changes its sign during the reflection (i.e.,
$\Delta_{zx}$ and $\Delta_x$) becomes zero as shown in
Figs.~\ref{fig:cur}(c) and \ref{fig:cur}(d). Correspondingly, the
other component is enhanced. When the surface is rough, both of the
components are strongly suppressed due to the random scatterings. 

\begin{figure}[tb]
	\includegraphics[width=0.48\textwidth]{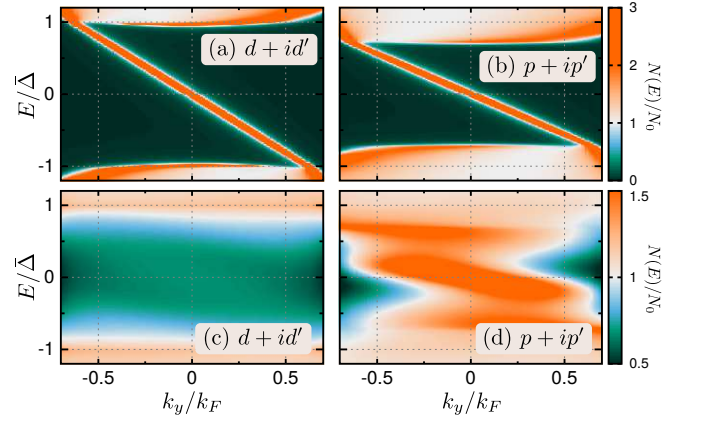}
	\caption{Angle-resolved density of states at $k_z/k_F =
	1/\sqrt{2}$ for the (a)(c) $d+id'$-wave and (b)(d) $p+ip'$-wave SCs. 
	The results are obtained at $x=0$ for the clean case [(a) and (b)]
	and at $x=w$ for the rough case with $\xi_0/\ell=0.5$ [(c) and (d)]. 
	The ARDOS are normalized to its value in the normal state. }
	\label{fig:ardos_a50}
\end{figure}

The difference in the robustness of the CSC comes from the symmetry of 
 the subdominant Cooper pairs induced by the local
inversion symmetry breaking at a surface. The inversion-symmetry
breaking results in the parity mixing of the pair amplitudes
\cite{Tanaka_07}. Namely, odd-parity (even-parity) pairings are induced at a
surface of the $d+id'$-wave ($p+ip'$-wave) SC. The $p_z$- and $s$-wave
pair amplitudes 
(i.e., subdominant pairs with the lowest azimuthal quantum number) in each SC are shown
in Figs.~\ref{fig:cur}(e) and \ref{fig:cur}(f), where we fix
$\omega_n=\omega_0$.  
The $s$-wave subdominant pairs plays an important role under an
disordered potential, whereas $p_z$-wave does not.  The $s$-wave pairs
$\langle f \rangle$ act as an effective pair potential in a 
disordered region [See Eq.~\eqref{eq:eta1}]. Namely, the disordered
region of the $p+ip'$-wave SC becomes an effective $s$-wave SC rather
than a normal metal. Consequently, the chiral current of the
$p+ip'$-wave SC flows along the internal interface at $x=w$.  In
Appendix, we show that $\check{g}$ at the internal interface is
qualitatively the same as that at a surface of a $p$-wave SC. The
chiral current does not flow at the internal interface in $d+id'$-wave
cases because the anisotropic $p_z$-wave pairs can not act as an
effective pair potential [i.e., $\langle f \rangle =0$ in Eq.~\eqref{eq:eta1}]. 

\begin{figure}[tb]
	\includegraphics[width=0.48\textwidth]{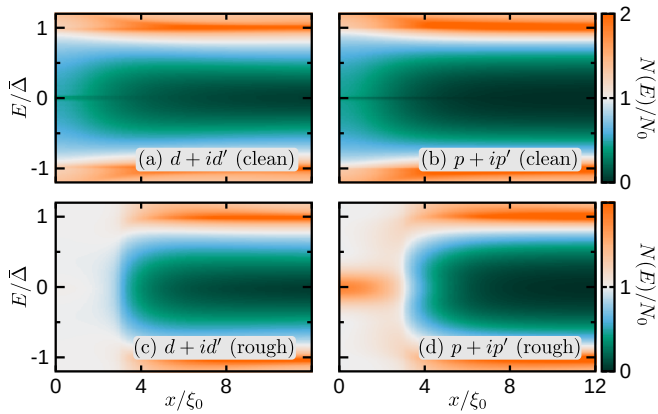}
	\caption{Local density of states of the (a)(c) $d+id'$-wave and
	(b)(d) $p+ip'$-wave SCs. The surface roughness is set to $\xi_0/\ell
	 = 0.5$ and $w=3\xi_0$ in (c) and (d). The LDOS at $x=0$ is enhanced because of the
	 chiral surface states. In $d+id'$-wave case, 
	 the disordered region can be regarded as a normal metal [i.e., $N(E)=N_0$]. 
	 The LDOS is normalized to its value in the normal state $N_0$.  }
	\label{fig:ldos_a50}
\end{figure}

The angle-resolved DOS (ARDOS) for the $d+id'$- and $p+ip'$-wave SCs
are compared in Fig.~\ref{fig:ardos_a50}, where we fix
$k_z=k_F/\sqrt{2}$. The ARDOS with $\rho=0$ ($\rho = 0.5$) are
obtained at the surface (internal interface). In the clean limit, the
chiral surface states are prominent in each SC. When the surface is
rough, the chiral states for the $d+id'$-wave SC vanishes [Fig.~\ref{fig:ardos_a50}(c)], 
whereas that for the $p+ip'$-wave SC is
robust [Fig.~\ref{fig:ardos_a50}(d)]. 
The LDOS can be calculated by integrating ARDOS. The results are shown
in Fig.~\ref{fig:ldos_a50}. We see the chiral surface states appear at
the surface; the LDOS increases at the surface (light blue region) as
shown in Figs.~\ref{fig:ldos_a50}(a) and \ref{fig:ldos_a50}(b). Under
the surface roughness, $N(x,E)=N_0$ in the disordered region of the 
$d+id'$-wave case, meaning that the disordered region becomes a normal
metal. In the $p+ip'$-wave SC, on the contrary, the LDOS has a peak
structure in the disordered region, reflecting the emergence of the effective 
$s$-wave superconductivity in the disordered region. To detect the
chiral surface states of the $d+id'$-wave SC, one has to pay close
attention to 
the surface quality because they are very sensitive to the roughness. 

\begin{figure}[tb]
	\includegraphics[width=0.48\textwidth]{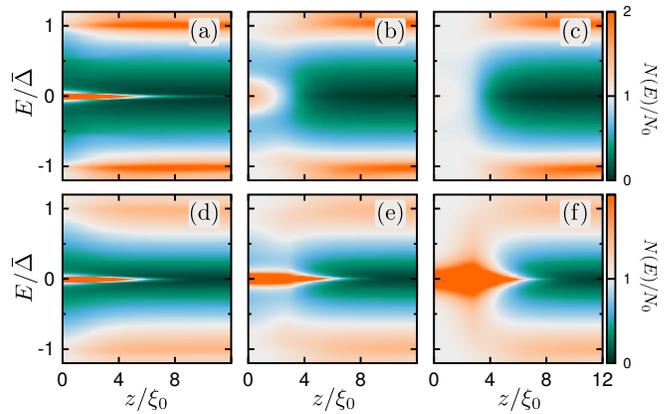}
	\caption{Effects of the surface roughness on the dispersion-less
	surface states of (a)-(c) the $d+id'$-wave and (d)-(f) $p_z$-wave
	SCs.
	The surface is perpendicular to the $z$ axis. 
	The strength of the roughness is set to 
	(a)(d) $\xi_0/\ell=0.0$, (b)(e) $0.2$, and (c)(f) $0.5$. 
	The results are obtained from
	the self-consistent $\Delta$ (not shown). The surface state in the
	$z$ direction of $d+id'$-wave case is much more fragile that those
	of $p_z$-wave case. }
	\label{fig:ldos_a00}
\end{figure}

\textit{Andreev bound states at $c$-axis surface}.---%
At the surface in the $c$-axis direction of the $d+id'$-wave SC, the
dispersion-less zero-energy states (ZESs) appear.\cite{kobayashi_15,tamura_17,suzuki_20} 
The effects of the
surface roughness are shown in Fig.~\ref{fig:ldos_a00}, where we also
show the results for a $p_z$-wave SC (i.e., polar state with
$\Delta_{\bs{k}} \sim p_z$) as a reference\cite{Hara}. 
The ZESs for both SCs are prominent in the clean limit
[Figs.~\ref{fig:ldos_a00}(a) and \ref{fig:ldos_a00}(d)].  However, in
the $d+id'$-wave case, the ZESs become broader even by weak surface
roughness (e.g., $\xi_0/\ell = 0.2$). 
Contrary to the $p_z$-wave SC \cite{Beenakker_PRB_2012, Suzuki_15, Ikegaya_16}, the ZESs
of the $d+id'$-wave SC disappear even for the weak disorder (i.e.,
$\xi_0 < \ell$). 

The fragility of the ZESs can be explained by the absence of the
$s$-wave subdominant pairs $\langle f \rangle$. 
The $p_z$-wave SC has robust ZESs supported by 
the $s$-wave pairs\cite{Suzuki_15} (i.e., effective pair
potential). On the other hand, the subdominant pairs for the $d+id'$-wave SC
are $p_x+ip_y$-wave-like pairs because of the phase winding at a fixed
$k_z$. Anisotropic $p_x +ip_y$-wave pairs do not act as an effective
pair potential [i.e., $\langle f \rangle = 0$ in Eq.~\eqref{eq:eta1}]. Therefore, the ZES at a
surface in the $c$-axis direction of a $d+id'$-wave SC are fragile
against roughness.

\textit{Discussion}.---%
The important factor determining the robustness of surface states
is only the presence of subdominant $s$-wave pairing induced at a
surface. Therefore, we can
generalize our knowledge to higher order chiral superconductors.
We have confirmed that the chiral surface states of
$f_{x(5z^2-1)}+if_{y(5z^2-1)}'$-wave SC are fragile against roughness
because of the absence of the $s$-wave pairs. Similarly, we can
anticipate fragile dispersion-less ZESs in the $f_{(x^2-y^2)z} +i
f_{xyz}$-wave SC since no $s$-wave pairing is expected. 

In this Letter, we employ the simple single-band model
and ignore the multi-orbital nature of SRO. The
fragility of the surface states is owing to the absence of 
$s$-wave Cooper pairs at the surface. 
In a $d+id$-wave SC, such $s$-wave subdominant pairs can be induced 
only in extreme cases: the scatterings by roughness cause 
a constructive interference for $s$-wave pairs. 
Therefore, the fragility of the surface states of
the $d+id'$-wave SC would be irrelevant to the details
of the model. Studying the roughness effects in detail
with more realistic three-orbital models~\cite{agterberg_20(r)} would
be an important future task, where the surface states would be suffered 
additionally from more complicated inter-band scatterings.

Here we briefly note that there are several experimental findings that
appears to contradict the $d+id'$-wave states in SRO.
For instance, a recent specific-heat measurement suggests the absence
of split between the onset temperatures for the superconductivity and
broken time-reversal symmetry state~\cite{mackenzie_21}, which seems
to contradict the two-component superconducting state.  Moreover, a
recent Josephson current measurement implies a
time-reversal invariant superconducting state~\cite{kashiwaya_19}.
Resolving such inconsistencies remains as an important
future task.

\textit{Conclusion}.---
We have investigated the effects of surface roughness on
the surface states of the ($d_{zx}+id_{yz}$)-wave SC.  Utilizing the
quasiclassical Eilenberger theory, we have demonstrated that the
surface states of the ($d_{zx}+id_{yz}$)-wave SC are easily destroyed
by surface roughness.  Since the surface roughness is inevitable
in real-life experiments, the absence of the experimental signatures
from the surface states~\cite{moler_05, nelson_07, flouquet_09,
kivelson_13, madhavan_20} would not be clearly inconsistent with the
inter-orbital ($d_{zx}+id_{yz}$)-wave superconducting state in SRO~\cite{agterberg_20(r)}.

\begin{acknowledgments}
We are grateful to A.~Brinkman, Y.~Asano, and 
T.~Kokkeler for the fruitful discussions. 
S.-I.~S. is supported by 
JSPS Postdoctoral Fellowship for Overseas Researchers
and a Grant-in-Aid for JSPS Fellows
(JSPS KAKENHI Grant No. JP19J02005),  
and thanks the University of Twente for  hospitality.
S.~I. is supported by a Grant-in-Aid
for JSPS Fellows (JSPS KAKENHI Grant No. JP21J00041).
\end{acknowledgments}

%

 
\appendix
\section{Effects of self-energy in a DN attached to a $p$-wave SC}
In this section, we consider a simplified theoretical model: the interface between
a dirty normal metal (DN) and a 
$p$-wave SC. For simplicity, we ignore the spatial dependence of the Green's functions
near the interface. The Riccati equations in the DN and SC are
\begin{align}
  \bs{v}_F \cdot \bs{\nabla}  \gamma_n
	+ 2 \tilde{\omega} \gamma_n
	  -     \eta 
		+ \ut{\eta}
	\gamma_n^2= 0, 
	\\
  \bs{v}_F \cdot \bs{\nabla}  \gamma_s
	+ 2 \omega \gamma_s
	  - \Delta_k
		+ \Delta_k^*
	\gamma_s^2= 0. 
\end{align}
with
\begin{align}
	& \tilde{\omega} = 
	\omega_n +\frac{\mathrm{Re} \langle g \rangle }{2 \tau_0}, 
	\hspace{6mm}
	\eta = \frac{\langle f \rangle}{2 \tau_0}, 
	\hspace{6mm}
	\ut{\eta} = - \frac{\langle f \rangle^*}{2 \tau_0}. 
\end{align}
At the interface, $\gamma$ can be obtained: 
\begin{align}
  &   \gamma _n 
	=-\frac{1}{\ut{\eta}}
	\left[ \tilde{\omega} - \sqrt{ \tilde{\omega}^2 + \eta \ut{\eta}} \right]
	, 
	\hspace{3mm}
  \gamma _s = \frac{\Delta  _k}{\omega_n + \Omega_n}, 
  	\notag
	\\
  &   
	\ut{\gamma}_n 
	= \frac{1}{\eta}
	\left[ \tilde{\omega} - \sqrt{ \tilde{\omega}^2 + \eta \ut{\eta}} \right]
	\hspace{3mm}
	\ut{\gamma}_s =-\frac{\Delta^*_k}{\omega_n + \Omega_n}, 
\label{}
\end{align}
where $\Omega_n = \sqrt{\omega_n^2 + |\Delta_k|^2}$. 
The normal Green's function, for example, can be obtained from them:
\begin{align}
	& g(+k,x=0,i \omega_n) = \frac
	{1+\gamma_n \ut{\gamma}_s(k)}
	{1-\gamma_n \ut{\gamma}_s(k)}, 
  \\
	& g(-k,x=0,i \omega_n) = \frac
	{1+\gamma_s(-k) \ut{\gamma}_n}
	{1-\gamma_s(-k) \ut{\gamma}_n}, 
	\label{eq:gF}
\end{align}
The Green's functions at the surface of a semi-infinite $p$-wave SC are calculated
from the coherence functions: 
\begin{align}
	g_{\mathrm{PW}}(\pm k,x=0,i \omega_n) = \frac
	{1+\gamma_s(-k) \ut{\gamma}_s(k)}
	{1-\gamma_s(-k) \ut{\gamma}_s(k)}. 
	\label{eq:gF2}
\end{align}
Comparing Eqs.~\eqref{eq:gF} and \eqref{eq:gF2}, we see the 
similarity when . Note that this similarity never appears 
in the $d$-, $f$- and $g$-wave SCs
because $\langle f \rangle = 0$ in those SCs.

\end{document}